\documentclass[cameraready]{Interspeech}
\usepackage{tabularx}  
\usepackage{booktabs}  
\usepackage{multirow}  
\usepackage{fix-cm}
\usepackage{amsmath}   
\usepackage{booktabs}  
\usepackage{graphicx}  
\usepackage{multirow}
\usepackage{color}
\usepackage{enumitem}
\usepackage{threeparttable}
\usepackage{subcaption}
\usepackage{amssymb}

\title{Lung-SRAD: Spectral-Aware Regularized Audio DASS with Dual-Axis Patch-Mix Contrastive Learning for Respiratory Sound Classification}

\usepackage[fontsize=9pt]{fontsize}

\author[affiliation={1}]{Hemansh}{Shridhar} 
\author[affiliation={1}]{Miika}{Toikkanen}
\author[affiliation={2,3}, orcid=0000-0003-0111-300X]{June-Woo}{Kim$^\dagger$}

\address{
    $^1$ RSC LAB, MODULABS, Republic of Korea \\
    $^2$ Department of Electronic Engineering, Wonkwang University, Republic of Korea \\
    $^3$ AI Convergence Research Institute, Wonkwang University, Republic of Korea
}

\email{shridharhemansh@gmail.com, kaen2891@wku.ac.kr}


\usepackage{comment}

\begin{document}

\maketitle
\renewcommand{\thefootnote}{$\dagger$}
\footnotetext{Corresponding author.}

\begin{abstract}
Recent respiratory sound classification (RSC) studies largely rely on CLS-token driven self-attention architectures such as the Audio Spectrogram Transformer (AST). While effective at modeling global context, recent analyses suggest a low-pass filtering behavior that may reduce sensitivity to localized abnormal patterns. In this work, we investigate State Space Models (SSMs) as an alternative backbone for RSC. Using the Distilled Audio State Space model, we analyze intermediate representations through spectral response curves and observe stronger preservation of mid-to-high spatial-frequency components. Based on these observations, we introduce spectral-aware layer regularization using Gaussian convolution applied to selected layers. We further propose Dual-Axis Patch-Mix contrastive learning tailored to SSM-based audio models for robust representation learning.    
Experiments on the ICBHI benchmark show that our approach achieves 64.48\% score, outperforming the AST baseline by 5\%. Code is available at \textcolor{cyan}{\href{https://github.com/RSC-Toolkit/Lung-SRAD}{https://github.com/RSC-Toolkit/Lung-SRAD}}.

\end{abstract}


\section{Introduction}\label{sec:introduction}


Abnormal lung sounds like ~\emph{crackles} and ~\emph{wheezes} are key indicators of respiratory disorders such as pneumonia, COPD, and asthma, which account for nearly four million deaths annually~\cite{grana2024assessing}.
Crackles are associated with diseases affecting the lung parenchyma and manifest as discontinuous acoustic events~\cite{flietstra2011automated}, whereas wheezes are characterized by continuous sounds that often appear as narrowband patterns in time–frequency representations~\cite{bohadana2014fundamentals}. 
These abnormal sounds manifest as short-duration, localized spectro-temporal structures, corresponding to rapid spatial variations across neighboring patches in a 2D spectrogram representation, analogous to \emph{high-frequency} components in image analysis rather than acoustic frequency along the spectrogram’s vertical axis. Because abnormal events are sparse and co-occur with dominant normal breathing patterns, preserving such localized variation is critical~\cite{takeishi2014anomaly}.

In respiratory sound classification (RSC), many recent high-performing methods rely on Transformer-based architectures, such as Audio Spectrogram Transformer (AST)~\cite{kim2024stethoscope, kim2025adaptive, gong2021ast, bae2023patch, kim2023adversarial, kim2024repaugment, kim2025tri}, Audio-CLAP~\cite{wu2023large, kim2024bts, toikkanen2025improving, koo2026empowering}, BEATs~\cite{chen2023beats, jeong2025patient}, or other variants~\cite{niizumi2024masked}, as the primary feature extraction backbone.
While these approaches achieve strong performance, they exhibit some limitations worth noting.
$(i)$ AST relies on global self-attention pooled into a single CLS token, and recent theory suggests a spectral limitation: softmax-based self-attention can behave as a \emph{low-pass filter} in the feature domain, progressively suppressing high-frequency inter-token variation while retaining dominant low-frequency components~\cite{wanganti}. This smoothing tendency may reduce sensitivity to localized abnormal events~\cite{zhangattention}. 
$(ii)$ The quadratic time and memory complexity of self-attention makes Transformer-based models computationally expensive for long sequences.
Motivated by these observations, we investigate audio state space models (SSMs) through their frequency response as an efficient alternative for preserving clinically relevant localized spectrogram patterns:


\begin{itemize}
    \item 
    We present the first application of a distilled State Space Model (DASS~\cite{bhati2024dass}) for RSC.
    When initialized with AudioSet-distilled weights, the model achieves 61.06\% Score on the ICBHI dataset~\cite{rocha2017alpha} using simple fine-tuning.
    
    \item 
    We analyze intermediate representations through spectral response curves, revealing how layers capture spatial-frequency components. Based on this, we propose a spectral-aware regularization strategy using Gaussian convolution applied to selected layers to improve RSC performance.

    \item 
    We introduce a Dual-Axis Patch-Mix supervised contrastive learning strategy designed for SSM-based audio models, achieving a Score of 64.48\% on the ICBHI dataset.

\end{itemize}

\begin{figure*}[t]
    \centering
    \includegraphics[width=\textwidth]{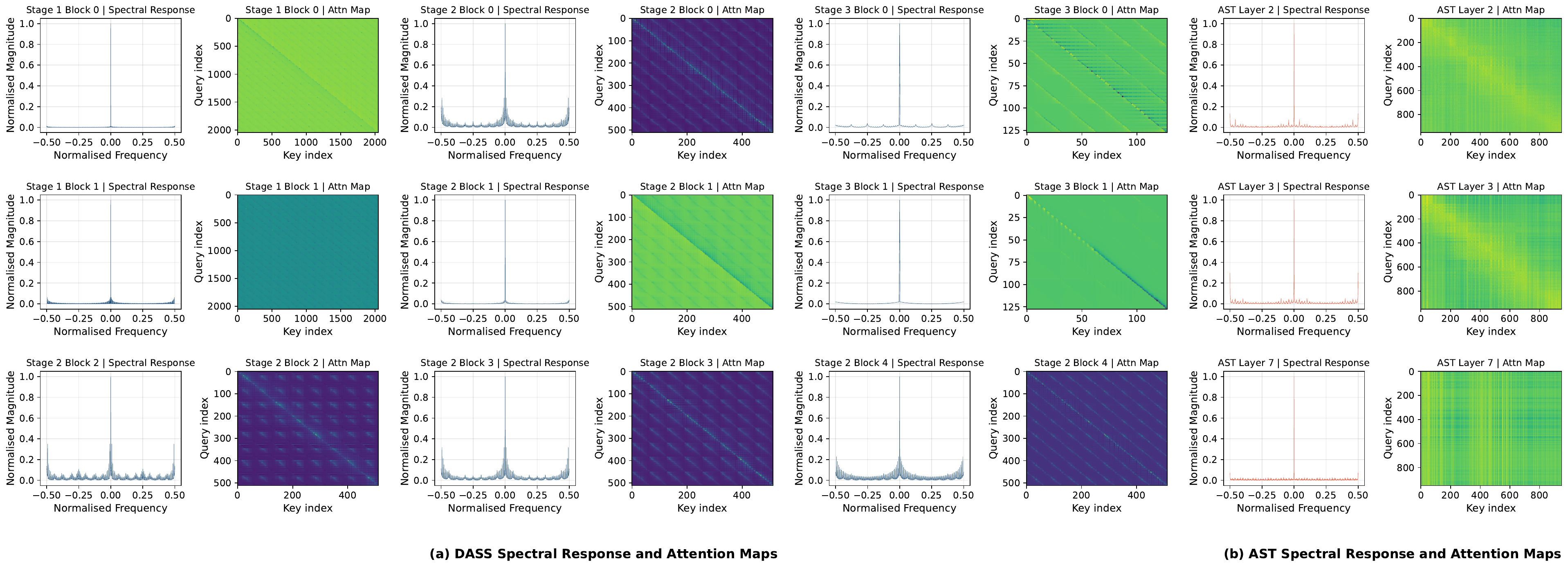}
    \caption{
    Spectral filter responses and attention maps of fine-tuned DASS (blue) and AST (red) on the ICBHI dataset. DASS Stage~2 layers (blocks) retain prominent mid-to-high frequency components while AST exhibits dominant low-pass component at all depths. 
}
    \vspace{-5mm}
    \label{fig:Spectral_Response}
    
\end{figure*}
\section{Preliminaries}
\label{sec:relatedwork}
\subsection{Dataset Description}
We use the ICBHI~\cite{rocha2017alpha} dataset, comprising 5.5 hours of recordings and 6,898 breathing cycles with four lung sounds: \emph{Normal, Crackle, Wheeze} and \emph{Crackle + Wheeze (Both)}. We follow the official 60\% train and 40\% test sets split, with no patient overlap between them, resulting in 4,142 cycles and 2,756 cycles. 

\subsection{Training Details}
All cycles were standardized to 8 seconds and resampled to 16 kHz~\cite{kim2024stethoscope, kim2025adaptive, bae2023patch, kim2023adversarial, kim2024repaugment, kim2024bts, kim2025tri}. SpecAugment~\cite{park19e_interspeech} is applied with a maximum mask length of 160 frames in time and 48 bins in frequency, without time-warping. We train all models using Adam~\cite{kingma2014adam} with a learning rate of \(5\times10^{-5}\) and batch size 16. 
Results are reported as the mean over five random seeds.







\subsection{Evaluation Metrics}
We evaluate using Sensitivity ($S_e$), Specificity ($S_p$), and the ICBHI Score~\cite{rocha2017alpha}, defined as 
$S_c = (S_e + S_p)/2$, following the official challenge protocol. 
$S_e$ measures the correct detection of abnormal respiratory sounds (\emph{crackle, wheeze, both}), whereas $S_p$ measures the correct identification of normal sounds.
\section{Methodology}
\label{sec:Methodology}
We introduce \textbf{Lung-SRAD} (Spectral-Aware Regularized Audio DASS for Lung Sounds) to address architectural limitations and spectral biases for RSC, with theoretical motivation and empirical validation for each design choice. 

\subsection{State Space Models (SSMs)}

\noindent\textbf{SSMs.} SSMs map an input sequence $x(t)\in\mathbb{R}^D$ to an output $y(t)$ through a latent state $h(t)\in\mathbb{R}^N$:
\begin{equation}\label{equ1}
    h'(t) = A h(t) + B x(t), \qquad y(t) = C h(t).
\end{equation}
where $A \in \mathbb{R}^{N \times N}$ is the state transition matrix 
that captures the system dynamics, $B \in \mathbb{R}^{N \times 1}$ and 
$C \in \mathbb{R}^{1 \times N}$ are input and output projection 
matrices, respectively. After zero-order hold discretization~\cite{gu2024mamba}, 
continuous parameters $(A, B)$ are mapped to discrete counterparts 
$(\bar{A}, \bar{B})$ through a learnable step size $\Delta$ via 
$\bar{A} = e^{\Delta A}$ and $\bar{B} = (\Delta A)^{-1}(e^{\Delta 
A} - I)\,\Delta B$, leading to:
\begin{equation}
    h_t = \bar{A} h_{t-1} + \bar{B} x_t, \qquad y_t = C h_t,
\end{equation}
Thus, SSMs combine recurrent modeling with convolutional efficiency while maintaining linear complexity.

\vspace{2mm}
\noindent\textbf{2D Selective State Space Scanning (SS2D). }
Mamba's~\cite{gu2024mamba} selective scan is defined for 1D sequences, whereas time--frequency representations are 2D. Following VMamba~\cite{liu2024vmamba}, \emph{2D Selective Scan (SS2D)} converts a 2D feature map into four directional sequences (row-wise, column-wise, and their reverses), applies selective SSM processing independently along each direction, and then reshapes the directional output back to the 2D grid. This design enables each location to integrate contextual information from multiple traversal directions while retaining the linear-time scanning property of selective SSMs.

Given a feature map $\mathbf{X} \in \mathbb{R}^{H \times W \times C}$, the four directional outputs are merged as:
\begin{equation}
    \mathbf{Y} = \mathrm{CrossMerge} \left( \mathbf{y}^{(1)}, \mathbf{y}^{(2)}, \mathbf{y}^{(3)}, \mathbf{y}^{(4)} \right) \in \mathbb{R}^{H \times W \times C},
\end{equation}
where $\mathbf{y}^{(k)}$ is the directional output obtained by applying a selective SSM along the $k$-th scan direction, thereby restoring the 2D structure and integrating contextual information from four traversal directions.

\subsection{DASS Architecture}

DASS~\cite{bhati2024dass} (Distilled Audio SSM) is a hierarchical audio state space model 
built upon the VMamba backbone for spectrogram-based audio classification. Given a mel-spectrogram $\mathbf{X} \in \mathbb{R}^{T \times F \times 1}$, it constructs a multi-scale feature hierarchy using four stages of SS2D blocks, where each stage is followed by patch-merging layers that reduce the time--frequency resolution while increasing the channel dimension. The final-stage features are summarized by global pooling and fed into a linear classifier.
DASS is initialized with AudioSet~\cite{audioset}-distilled weights transferred from an ensemble of Transformer teachers (AST~\cite{gong2021ast} and HTS-AT~\cite{chen2022hts}). 
We fine-tune these pretrained weights for RSC.

\subsection{Architectural Gap and Spectral Filtering Behavior}
AST~\cite{gong2021ast} performs global self-attention and uses a \texttt{CLS} token for classification. 
Wang et al.~\cite{wanganti} show that softmax attention acts as a \emph{low-pass filter}, suppressing mid to high spatial frequency inter-token variation while preserving low-frequency global structure. 
This observation is particularly relevant for RSC, where abnormal events such as crackles and wheezes often exhibit more rapidly varying time--frequency patterns than normal breathing. 

To compare the spectral characteristics of AST and DASS, we examine their intermediate layer filter responses using the analytical framework of~\cite{wanganti}. 
Here, \emph{frequency} refers to \emph{spatial-feature variation}, i.e., the variation rate across 2D spectrogram patch tokens in the learned feature space. 
For a learned weight matrix $\mathbf{M}$ (the attention matrix in AST or the SS2D mixing matrix in DASS), we compute its Fourier-domain response $\boldsymbol{\Lambda} = \mathcal{F}\mathbf{M}\mathcal{F}^{-1}$, where $\mathcal{F}$ is the DFT matrix with basis vectors $\mathbf{f}_k = \frac{1}{\sqrt{n}}\bigl[e^{2\pi j(k-1)\cdot 0}, \dots, e^{2\pi j(k-1)\cdot(n-1)}\bigr]^{\top}$. 
The spectral response intensity at the $i$-th frequency band is evaluated as $\frac{1}{n}\sum_j|\Lambda_{ij}|$. 
Lower responses at higher bands indicate low-pass behavior that suppresses fine-grained inter-token variation.

As in Fig.~\ref{fig:Spectral_Response}, AST exhibits a predominantly low-pass profile across layers, consistent with~\cite{wanganti}. Moreover, deeper layers increasingly concentrate attention on the \texttt{CLS} token, aligning with the \emph{attention sink} phenomenon~\cite{xiaoefficient} and yielding more uniform token representations.
Since only the \texttt{CLS} token embedding is directly optimized for classification, this bottleneck may encourage reliance on dataset-specific shortcut cues~\cite{kim2024stethoscope}.

In contrast, DASS replaces attention with SS2D, which aggregates features from the full time--frequency feature map via spatial pooling, allowing the classifier to utilize distributed representations rather than a single token.
As observed in Fig.~\ref{fig:Spectral_Response}, DASS retains significant spectral magnitude across multiple mid to high frequency bands, in contrast to AST which concentrates energy at the DC component with negligible 
magnitude at mid to high frequencies. Fig.~\ref{fig:Spectral_Response} further shows stage-wise diversity in DASS: early and late stages emphasize low-frequency components capturing global context, whereas intermediate stages show stronger mid-to-high frequency responses. 
This pattern can improve the recognition of short-duration abnormal respiratory events. 

\begin{figure}[t]
\centering
\includegraphics[width=\columnwidth]{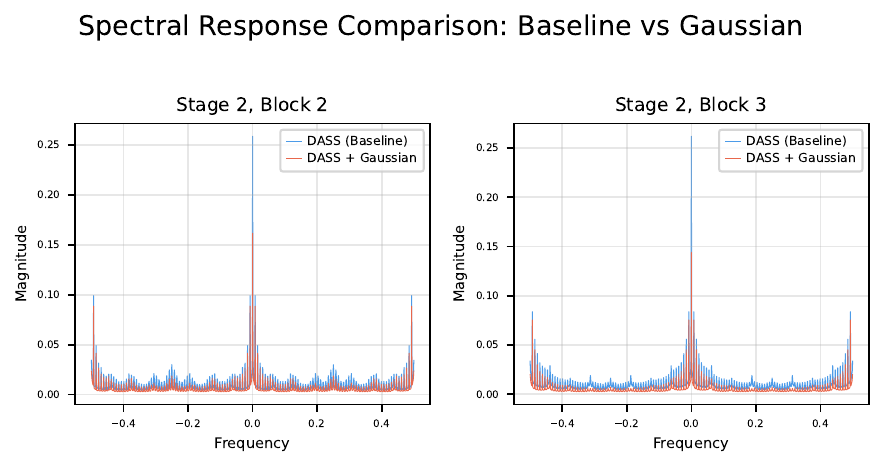}
\caption{Spectral responses for DASS baseline (blue) and DASS with Gaussian smoothing (red) for Stage~2 Blocks~2--3. Gaussian smoothing attenuates dominant spectral peaks while preserving the harmonic structure of the frequency response.
}
\label{fig:Spectral_comparison}
\vspace{-5mm}
\end{figure}

\begin{table*}[!t]
    \centering
    \caption{Overall RSC results on the ICBHI datasets (60--40\% official). In the pretraining data column, IN, AS, and LA denote ImageNet~\cite{deng2009imagenet}, AudioSet~\cite{audioset}, and LAION-Audio-630K~\cite{wu2023large}, respectively. $*$ denotes the previous state-of-the-art ICBHI Score. \textbf{Best} and \underline{second-best} results. Note that our performance on the 2-class is derived from 4-class model training weights, indicating the same $S_p$.}
    
    \label{table1}
    \renewcommand{\arraystretch}{1}
    \addtolength{\tabcolsep}{8pt}
    \resizebox{\linewidth}{!}{
    \begin{tabular}{p{2pt}llcllll}
    \toprule
    & Method & Backbone & Pretraining Data & Venue & $S_p$\,(\%) & $S_e$\,(\%) & \textbf{Score}\,(\%) \\
    \hline \midrule


    
    
    
    \multirow{12.5}{*}{\rotatebox[origin=c]{90}{\textbf{4-class eval.}}} & Bae \textit{et al.} \cite{bae2023patch}\, (Fine-tuning)  & AST & IN\,+\,AS & \textit{INTERSPEECH`23} & $\text{77.14}$ & $\text{41.97}$ & $\text{59.55}$ \\
    
    & Bae \textit{et al.} \cite{bae2023patch}\, (Patch-Mix CL) & AST & IN\,+\,AS & \textit{INTERSPEECH`23} & $\text{81.66}$ & $\text{{43.07}}$ & $\text{62.37}$ \\
    
    & Kim \textit{et al.} \cite{kim2024stethoscope}\, (SG-SCL) & AST & IN\,+\,AS & \textit{ICASSP`24} & $\text{{79.87}}$ & $\text{{43.55}}$ & $\text{{61.71}}$ \\

    & Xiao \textit{et al.} 
    \cite{xiao24_interspeech} (LungAdapter) & \text{AST} & IN\,+\,AS & \textit{INTERSPEECH`24} & $\text{{80.43}}$ & $\text{44.37}$ & $\text{{62.40}}$ \\

    
    & Kim \textit{et al.} 
    \cite{kim2024bts} (BTS) & CLAP & LA & \textit{INTERSPEECH`24} & $\text{81.40}$ & $\text{{45.67}}$ & $\text{{63.54}}$ \\


    & Jeong \textit{et al.} \cite{jeong2025patient} (PAFA) & \text{BEATs} & AS & \textit{INTERSPEECH`25} & $\underline{{82.05}}$ & $\text{47.63}$ & $\underline{{64.84}}$ \\

    & Toikkanen \textit{et al.} \cite{toikkanen2025improving} (BTS++[$k = 30$]) & \text{CLAP} & LA & \textit{INTERSPEECH`25} & $\textbf{{89.49}}$ & $\text{41.89}$ & $\textbf{{65.69}}^\textbf{*}$ \\
    \cmidrule{2-8}
    
    & \textbf{DASS (Fine-tuning) [ours]} & \text{DASS} & AS & \textit{--} & $\text{{74.68}}_{\pm 2.50}$ & $\text{47.43}_{\pm 1.58}$ & $\text{61.06}_{\pm 1.27}$ \\

    & \textbf{DASS (Spectral-Aware Regularization) [ours]} & \text{DASS} & AS & \textit{--} & $\text{\text{76.72}}_{\pm 5.67}$ & $\underline{47.72}_{\pm 3.23}$ & $\text{62.22}_{\pm 1.29}$ \\

    & \textbf{DASS (Lung-SRAD) [ours]} & \text{DASS} & AS & \textit{--} & $\text{\text{79.53}}_{\pm 0.95}$ & $\textbf{49.42}_{\pm 1.19}$ & $\text{64.48}_{\pm 0.25}$ \\
    
    \midrule
    
    \multirow{8.5}{*}{\rotatebox[origin=c]{90}{\textbf{2-class eval.}}} & Bae \textit{et al.} \cite{bae2023patch}\, (Fine-tuning) & AST & IN\,+\,AS & \textit{INTERSPEECH`23} & $\text{77.14}$ & $\text{56.40}$ & $\text{66.77}$ \\
    
    
    & Bae \textit{et al.} \cite{bae2023patch}\, (Patch-Mix CL) & AST & IN\,+\,AS & \textit{INTERSPEECH`23} & $\text{\bf 81.66}$ & $\text{55.77}$ & $\text{68.71}$ \\
    & Kim \textit{et al.} \cite{kim2024stethoscope}\, (SG-SCL) & AST & IN\,+\,AS & \textit{ICASSP`24} & $\text{\underline{79.87}}$ & $\text{57.97}$ & $\text{68.93}$ \\
    & Jeong \textit{et al.} \cite{jeong2025patient} (PAFA) & \text{BEATs} & AS & \textit{INTERSPEECH`25} & $\text{{74.87}}$ & $\textbf{68.29}$ & $\underline{{72.08}}^\textbf{*}$ \\
    \cmidrule{2-8}
    & \textbf{DASS (Fine-tuning) [ours]} & \text{DASS} & AS & \textit{--} & $\text{{73.74}}_{\pm 3.38}$ & $\text{61.72}_{\pm 2.94}$ & $\text{68.20}_{\pm 2.05}$ \\

    & \textbf{DASS (Spectral-Aware Regularization) [ours]} & \text{DASS} & AS & \textit{--} & $\text{\text{76.72}}_{\pm 5.67}$ & $\text{60.29}_{\pm 3.96}$ & $\text{68.40}_{\pm 1.04}$ \\

    & \textbf{DASS (Lung-SRAD) [ours]} & \text{DASS} & AS & \textit{--} & $\text{\text{79.53}}_{\pm 0.95}$ & $\underline{65.61}_{\pm 0.24}$ & $\textbf{72.57}_{\pm 0.47}$ \\


    
    \bottomrule
    \end{tabular}}
    \vspace{-5mm}
\end{table*}

\subsection{Spectral-Aware Layer Regularization}\label{sec3.4}
While mid-to-high frequency responses can help capture abnormal respiratory events, overly strong local variations may bias the model toward fine-grained local patterns and weaken global contextual modeling. To control this effect, we apply Gaussian smoothing \emph{selectively} to the layers that show excessive mid-to-high frequency responses in the filter-response curves. We define a 1D Gaussian kernel as:
\begin{equation}
    G(k) = \frac{1}{\sqrt{2\pi\sigma^2}} 
    \exp\Bigl(-\frac{k^2}{2\sigma^2}\Bigr),
    \label{eq:gaussian_kernel}
\end{equation}
where $k \in \{-\lfloor\tfrac{K-1}{2}\rfloor, \dots, \lfloor\tfrac{K-1}{2}\rfloor\}$, $K$ is the kernel size and the kernel is $\ell_1$-normalized as $\tilde{G}(k) = G(k)/\sum_j G(j)$. Given an intermediate activation map $\mathbf{O}^{(l)} \in \mathbb{R}^{B \times C \times H \times W}$, where $l$ denotes the layer index, smoothing is applied depthwise and separably along the two spatial axes:
\begin{equation}
    \mathbf{O}^{(l)}_{\text{low}} = 
    \tilde{G}_H * (\tilde{G}_W * \mathbf{O}^{(l)}),
    \label{eq:separable_blur}
\end{equation}
where $\tilde{G}_H$ and $\tilde{G}_W$ are 1D Gaussian kernels applied depthwise along $H$ and $W$, respectively. In our experiments, we apply this operation to Stage~2 layers that exhibit prominent mid-to-high frequency peaks (e.g., Block~2 and Block~3). Consequently, these two blocks are selected for the smoothing operation in our experiments. 
As shown in Fig.~\ref{fig:Spectral_comparison}, Gaussian smoothing reduces the magnitude of these peaks while preserving the overall spectral trend. Empirically, this regularization improves $S_p$ while comparable $S_e$, as described in Table~\ref{table1}. 


\subsection{Dual-Axis Patch-Mix Contrastive Learning}\label{sec3.5}
Patch-Mix contrastive learning~\cite{bae2023patch} was originally proposed for the AST model, where spectrogram patches are treated as an unordered token set processed via self-attention. Conversely, VMamba processes inputs as structured 2D grids with directional state transitions along temporal and frequency scan paths. Naive patch mixing can disrupt the sequential state evolution, leading to unstable gains. To align contrastive perturbations with VMamba’s multi-directional scanning, we define axis-aligned patch mixing and corresponding contrastive losses.

\subsubsection{Axis-Aligned Patch Mixing}
Given a batch of spectrograms $\mathbf{X} \in \mathbb{R}^{B \times H \times W \times C}$, where $H, W$ denote the frequency and time axes, respectively, and $C$ is the embedding dimension after patch embedding. Let $\pi$ denote a random permutation over the batch dimension. Unlike AST, we avoid random patch selection over the flattened 2D grid. Instead, we replace consecutive segments, preserving structured state transitions along the orthogonal axis. 

\noindent\textbf{Temporal Patch-Mix.} For temporal alignment, we replace a consecutive temporal segment of width $w$:
\begin{equation}
\tilde{\mathbf{X}}_{:, :, t:t+w, :}^{\text{time}} = \mathbf{X}_{\pi, :, t:t+w, :}.
\end{equation}
Here, the segment width $w$ is determined by a mixing ratio $\lambda_t \sim \operatorname{Beta}(\beta, \beta)$ such that $w = \lfloor W(1-\lambda_t) \rfloor$, preserving frequency-wise state continuity while regularizing temporal state transitions along VMamba's horizontal scan paths.

\noindent\textbf{Frequency Patch-Mix.} Similarly, we replace a consecutive frequency band of height $h = \lfloor H(1-\lambda_f) \rfloor$:
\begin{equation}
\tilde{\mathbf{X}}_{:, f:f+h, :, :}^{\text{freq}} = \mathbf{X}_{\pi, f:f+h, :, :}.
\end{equation}
In this case, temporal continuity is preserved while patch-mixing is applied along the spectral dimension, aligning with the vertical scan paths. 

\subsubsection{Dual-Axis Contrastive Learning}
To regularize the model without compromising the backbone's pretrained capabilities, we use an asymmetric gradient strategy. 
Let $f_\theta(\cdot)$ be the VMamba backbone and $h(\cdot)$ be a 768-dimensional projection head. We define the normalized embeddings for the original input $\mathbf{X}_i$ and its mixed version $\tilde{\mathbf{X}}_i$ as:
\begin{equation}
\hat{\mathbf{q}}_i = \frac{h(f_\theta(\mathbf{X}_i))}{\left\lVert h(f_\theta(\mathbf{X}_i)) \right\rVert_2}, \quad 
\hat{\tilde{\mathbf{q}}}_i = \frac{h(\operatorname{sg}(f_\theta(\tilde{\mathbf{X}}_i)))}{\left\lVert h(\operatorname{sg}(f_\theta(\tilde{\mathbf{X}}_i))) \right\rVert_2}.
\end{equation}
Here, $\operatorname{sg}(\cdot)$ denotes the stop-gradient operation. By blocking gradients from the mixed samples, the contrastive objective encourages the projection space to be invariant to axis-aligned patch-mixing while preventing the distorted signals from collapsing the structured state-space dynamics of the backbone.

\noindent\textbf{Patch-Mix InfoNCE.} For a given axis $a \in \{\text{time, freq}\}$, the patch-mix contrastive loss $\mathcal{L}_{PM}^a$ is defined as:
\begin{equation}
\mathcal{L}_{PM}^a = -\sum_{i=1}^B \log \frac{\lambda_a e^{\text{sim}(\hat{\mathbf{q}}_i, \hat{\tilde{\mathbf{q}}}_i)} + (1-\lambda_a) e^{\text{sim}(\hat{\mathbf{q}}_i, \hat{\tilde{\mathbf{q}}}_{\pi(i)})}}{\sum_{j=1}^B e^{\text{sim}(\hat{\mathbf{q}}_i, \hat{\tilde{\mathbf{q}}}_j)}},
\end{equation}
where $\text{sim}(\mathbf{u}, \mathbf{v}) = \mathbf{u}^\top \mathbf{v} / \tau$ denotes the cosine similarity scaled by a temperature $\tau$. In our experiments, we set $\tau = 0.2$. 
The mixing ratio $\lambda_a$ is sampled from a symmetric Beta distribution, $\lambda_a \sim \operatorname{Beta}(\beta, \beta)$ with $\beta=1.0$, for each mini-batch.

\subsection{Total Objective: Lung-SRAD}
To fully leverage the directional scan paths of VMamba, we combine the supervised cross-entropy loss $\mathcal{L}_{CE}$ with dual-axis patch-mix supervised contrastive loss:
\begin{equation}
\mathcal{L}_{total} = \mathcal{L}_{CE} + \mathcal{L}_{PM}^{\text{time}} + \mathcal{L}_{PM}^{\text{freq}}.
\end{equation}
\section{Results}\label{sec:experiments}
\subsection{Main Results}
\noindent\textbf{4-Class Results.} Table~\ref{table1} summarizes the performance on the ICBHI benchmark under the official 60--40 patient independent split. 
With simple fine-tuning from AudioSet-distilled DASS initialization, the model achieves 61.06\% Score. 
Applying spectral-aware regularization with Gaussian convolution on selected intermediate layers improves performance to 62.22\% Score, primarily by increasing specificity ($S_p$ 74.68\% $\rightarrow$ 76.72\%) while maintaining sensitivity ($S_e$ 47.43\% $\rightarrow$ $47.72$\%). 
Finally, our Lung-SRAD further boosts the Score to 64.48\%, outperforming the AST baseline by 5\%. 

\noindent\textbf{2-Class Results.}
We further evaluate in the 2-class setting (normal vs. abnormal). Notably, the 2-class results are derived from the 4-class trained weights, indicating the same specificity across settings. DASS fine-tuning achieves 68.20\% Score, and spectral-aware regularization slightly improves to 68.40\% Score. With Lung-SRAD, the model reaches 72.57\% Score, exceeding the prior reported 2-class best Score in Table~\ref{table1}.

\begin{table}[t]
\centering
\caption{Impact of spectral-aware regularization and Patch-Mix design. \textbf{Best} and \underline{second-best} results. 
}
\label{tab2:impact_ablation_icbhi}
\setlength{\tabcolsep}{6pt}
\renewcommand{\arraystretch}{1}
\resizebox{\linewidth}{!}{
\begin{tabular}{lccc}
\toprule
Method & $S_p$\,(\%) & $S_e$\,(\%) & \textbf{Score} \,(\%) \\
\midrule
(1). DASS fine-tuning
& $74.68_{\,\pm 2.50}$ & $47.43_{\,\pm 1.58}$ & $61.06_{\,\pm 1.27}$ \\
(2). (1) + Spectral-Aware Regularization
& $76.72_{\,\pm 5.67}$ & $47.72_{\,\pm 3.23}$ & $62.22_{\,\pm 1.29}$ \\
(3). (2) + AST-style Patch-Mix CL
& $74.39_{\,\pm 3.93}$ & $50.46_{\,\pm 3.86}$ & $62.42_{\,\pm 0.33}$ \\
(4). (2) + Freq-only Patch-Mix CL
& $76.43_{\,\pm 6.23}$ & $49.94_{\,\pm 3.65}$ & $63.18_{\,\pm 1.31}$ \\
(5). (2) + Time-only Patch-Mix CL
& $76.68_{\,\pm 4.52}$ & $48.97_{\,\pm 3.52}$ & $62.83_{\,\pm 0.69}$ \\
(6). (2) + (4) + (5) Lung-SRAD 
& $\mathbf{79.53}_{\,\pm 0.95}$ & $49.42_{\,\pm 1.19}$ & $\mathbf{64.48}_{\,\pm 0.25}$ \\
\bottomrule
\end{tabular}
}
\vspace{-5mm}

\end{table}

\subsection{Impact of Regularization and Patch-Mix Design}
Table~\ref{tab2:impact_ablation_icbhi} presents a comparative study on the ICBHI 4-class setting to quantify the impact of each component. Starting from DASS fine-tuning, spectral-aware regularization improves the Score from 61.06\% $\rightarrow$ 62.22\%, primarily through higher specificity, which is consistent with reduced false positives. As analyzed in Section~\ref{sec3.4} using spectral response curves, intermediate layers increasingly emphasize higher-frequency components, which can be amplified during fine-tuning and contribute to false positives. Our spectral-aware regularization addresses these components, improving specificity (74.68\% $\rightarrow$ 76.72\%) while preserving sensitivity (47.43\% $\rightarrow$ 47.72\%).

Based on spectral-aware regularization, Patch-Mix supervised contrastive learning further improves performance. Consistent with our discussion in Section~\ref{sec3.5}, DASS benefits more from task-aligned Patch-Mix designs than the AST-style variant: frequency-only and time-only mixing improve the Score to 62.22\% $\rightarrow$ 63.18\% and 62.22\% $\rightarrow$ 62.83\%, respectively, while the proposed Dual-Axis Patch-Mix (Lung-SRAD) achieves the best overall Score of 64.48\%, indicating that jointly mixing both axes provides a stronger regularization than either axis alone.

\begin{figure}[t]
\centering
\includegraphics[width=\linewidth]{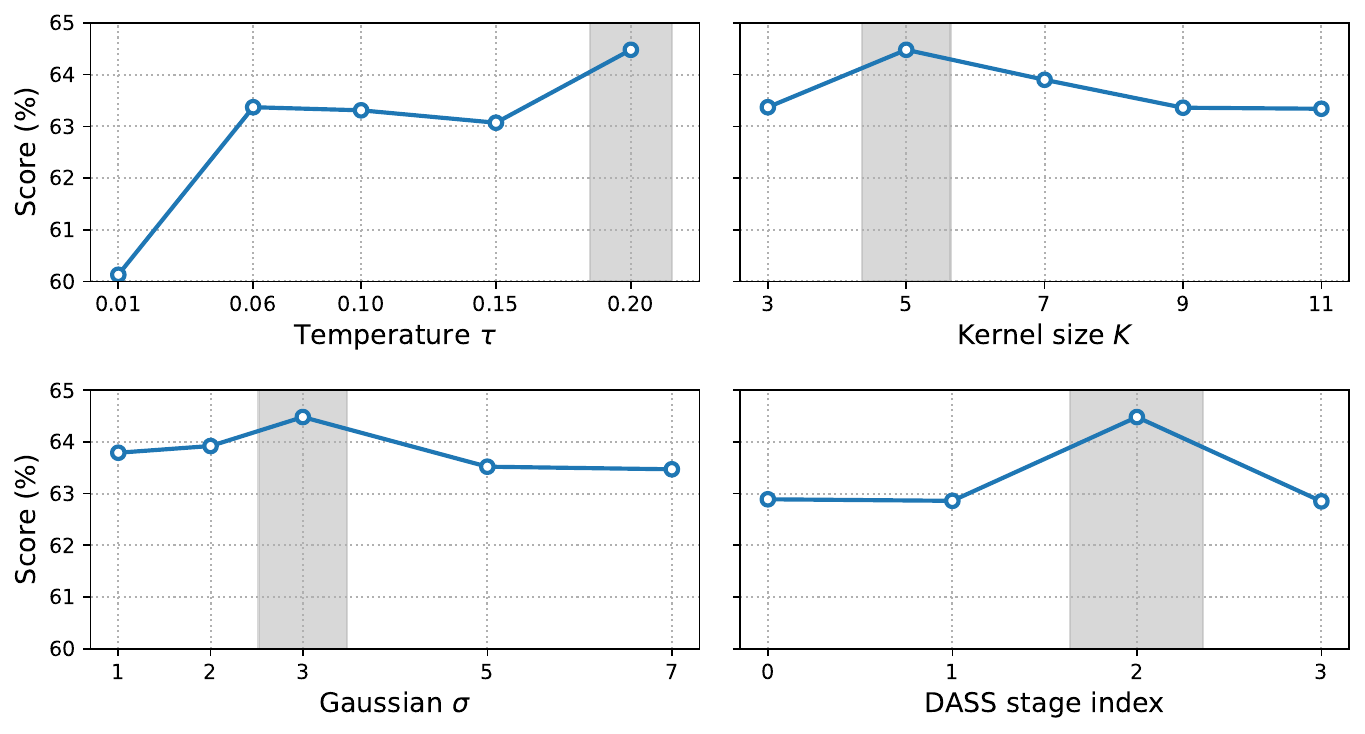}
\vspace{-3mm}
\caption{Hyperparameter analysis of Patch-Mix CL and spectral-aware regularization, evaluated by ICBHI Score.
}
\label{fig:hparam_ablation_2x2}
\vspace{-5mm}
\end{figure}

\subsection{Hyperparameter Analysis}
Fig.~\ref{fig:hparam_ablation_2x2} shows sensitivity to key hyperparameters. For Patch-Mix supervised contrastive learning, performance improves with temperature, peaking at $\tau=0.20$, while $\tau=0.01$ causes a large drop. For spectral-aware regularization, $K=5$ and $\sigma=3$ perform best, whereas larger $K$ or $\sigma$ reduce the Score. Applying the regularization at DASS stage 2 yields the largest gain, consistent with Fig.~\ref{fig:Spectral_Response} where stage 2 shows the strongest high-frequency emphasis in the spectral response curves.


\section{Conclusion}
\label{sec:conclusion}

In this work, we explored SSM as a backbone for RSC. We observed that the DASS architecture preserves mid-to-high spatial-frequency components important for capturing localized abnormal respiratory events. Based on this observation, we introduced spectral-aware regularization using Gaussian smoothing on selected intermediate layers, along with Dual-Axis Patch-Mix contrastive learning aligned with VMamba’s directional scanning. 
Our method achieved Scores of 64.48\% and 72.57\% on the ICBHI dataset in the 4- and 2-class settings, respectively, outperforming a strong AST baseline while maintaining a favorable balance between sensitivity and specificity.


\section{Acknowledgement}
This research was supported by the Regional Innovation System \& Education(RISE) program through the Jeonbuk RISE Center, funded by the Ministry of Education(MOE) and the Jeonbuk State, Republic of Korea(2026-RISE-13-WKU), and by the National Research Foundation of Korea(NRF) grant funded by the Korea government(MSIT) (grant no. RS-2025-16066662).

\section{Generative AI Use Disclosure}
Generative AI (ChatGPT) was used solely for grammar correction and linguistic polishing of this manuscript. The authors have verified all technical content and maintain full accountability for the work.

\bibliographystyle{IEEEtran}
\bibliography{mybib}
\end{document}